# Adsorption of Hydrogen in Graphene without Band Gap Opening at the Dirac Point

J M. García-Lastra[1,2]

[1]*Nano-Bio Spectroscopy group and ETSF Scientific Development Centre, Dpto. Física de Materiales, Universidad del País Vasco, Av. Tolosa 72, E-20018 San Sebastián, Spain*

[2]*Center for Atomic-scale Materials Design, Department of Physics, Technical University of Denmark, DK-2800 Kgs. Lyngby, Denmark*

**Abstract**

The band gap of periodically-doped graphene with hydrogen is investigated. It is found through a tight-binding model (TB) that for certain periodicities, called here NGPs (non-gap periodicities), no gap is opened at the Dirac point. This result is confirmed by Density Functional Theory (DFT) calculations. DFT results show that a tiny gap is opened for NGPs due to exchange effects, not taken into account in the TB model. However, this tiny gap is one or two orders of magnitude smaller than the gap opened for other periodicities different from NGPs. This finding opens up a new path for band gap engineering experiments in graphene.

The unusual semimetallic behavior of graphene, a single layer of graphite, was first studied theoretically by Wallace[1] and its recent synthesis[2] has renewed the interest for this material[3]. The great variety of physical and chemical properties derived from the electronic structure of graphene makes it very attractive for a range of applications, from catalysis[4] to nanoscale electronics[5]. It is remarkable that graphene is a gapless semiconductor, with no gap at the Dirac point[3] (the K and K' points in Fig. 1b). However, it is well known that the presence of defects such as vacancies or impurities[6-7], the interaction with a substrate[8], or the edge structure (in graphene nanoribbons)[9-11] can lead to a band gap opening in graphene[3].

In particular, the hydrogenation of graphene and its effect in the band gap has been investigated in several recent works[9, 11-17]. Combined Scanning Tunneling Microscopy (STM) and Angle-Resolved Photoemission Spectroscopy (ARPES) experiments showed that the patterned adsorption of atomic hydrogen onto the Moiré superlattice positions of graphene grown on an Ir(111) substrate produces a band gap opening in graphene[12]. Density Functional Theory (DFT) calculations reported in the literature[14, 17-18] also predict a band gap opening, which depends on the periodicity of the hydrogen adsorption, namely: i) In graphane, the fully hydrogenated graphene, the gap at the $\Gamma$ point (see Fig. 1b) is 3.5 eV, whereas the gap opening at the Dirac point is above 10 eV.[18] ii) In half-hydrogenated graphene the maximum of the valence band is at the Dirac point and the minimum of the conduction band is at the $\Gamma$ point, with a gap between them of 0.43 eV. The direct gap at the Dirac point is 1.16 eV[17]. iii) In a graphene sheet with a hydrogen atom adsorbed with a $4\vec{a}_1 \times 4\vec{a}_2$ periodicity, being $\vec{a}_1$ and $\vec{a}_2$ the basis vectors of the hexagonal lattice of graphene (see Fig. 1a), the gap opened at the Dirac point, which in this case is the gap of the material, is 1.25 eV[14].

After these observations the following question arises: does the periodical adsorption of hydrogen atoms in graphene lead to an opening of its gap for any periodicity? In order to answer this question we will first examine the main features of the structure of graphene and its reciprocal lattice, depicted in Fig. 1. A graphene sheet consists of a 2D hexagonal lattice with two atoms per unit cell (A and B atoms in figure 1a). The $p_z$ orbitals of the carbon atoms, which cannot overlap with $p_x$, $p_y$ or s orbitals (it is forbidden by symmetry), form π-bands. We can understand the band structure of graphene through a tight-binding (TB) model, making linear combinations of the $p_z$ orbitals of A and B atoms, denoted here as $p_z^A$ and $p_z^B$, respectively. The corresponding TB Hamiltonian for the unit cell of graphene, assuming the overlap between $p_z^A$ and $p_z^B$ orbitals to be negligible, can be written as:

$$\begin{pmatrix} \varepsilon_0 & -\gamma_0 f(\vec{k}) \\ -\gamma_0 f^*(\vec{k}) & \varepsilon_0 \end{pmatrix} \begin{pmatrix} c_A(\vec{k}) \\ c_B(\vec{k}) \end{pmatrix} = E(\vec{k}) \begin{pmatrix} c_A(\vec{k}) \\ c_B(\vec{k}) \end{pmatrix} \quad (1)$$

Where $\gamma_0$ is the transfer integral between first-neighbor $p_z$ orbitals (a typical value for it is 2.9 eV[19]), $\varepsilon_0$ is the energy reference from the $p_z$ orbital of a carbon atom (which in the following will be set to $\varepsilon_0 = 0$), and the phase factor is $f(\vec{k}) = \sum_{\vec{\delta}_j=1}^{3} e^{i\vec{k}\cdot\vec{\delta}_j}$, being $\vec{\delta}_j$ the nearest-neighbors vectors (see Fig. 1). $c_A(\vec{k})$ and $c_B(\vec{k})$ are respectively the coefficients of $p_z^A$ and $p_z^B$ orbitals in the linear combinations of the eigenfunctions. It can be noticed that at $\vec{k} = $ K, K', i.e. at the Dirac points, $f(\vec{k}) = 0$. Thus at the Dirac point, $p_z^A$ and $p_z^B$ orbitals are decoupled and degenerate in energy, leading to a zero gap semiconductor band structure.

The same TB scheme can be used to model a $m\cdot\vec{a}_1 \times n\cdot\vec{a}_2$ graphene supercell (SC), instead of the $\vec{a}_1 \times \vec{a}_2$ unit cell used in Eq.(1). The TB Hamiltonian of a $m\cdot\vec{a}_1 \times n\cdot\vec{a}_2$ SC at the Dirac point is a Hermitian matrix full of zeros, except for the coupling matrix elements between two nearest neighbors. These coupling matrix elements for two atoms connected by $\vec{\delta}_1$, $\vec{\delta}_2$ and $\vec{\delta}_3$ are $-\gamma_0$, $-\gamma_0 \cdot e^{2\pi i/3}$ and $-\gamma_0 \cdot e^{-2\pi i/3}$ (and their respective conjugated), respectively. For instance, the corresponding TB Hamiltonian for a $2\cdot\vec{a}_1 \times 2\cdot\vec{a}_2$ SC is:

$$-\gamma_0 \begin{pmatrix} 0 & 1 & 0 & e^{i2\pi/3} & 0 & e^{-i2\pi/3} & 0 & 0 \\ 1 & 0 & e^{-i2\pi/3} & 0 & e^{i2\pi/3} & 0 & 0 & 0 \\ 0 & e^{i2\pi/3} & 0 & 1 & 0 & 0 & 0 & e^{-i2\pi/3} \\ e^{-i2\pi/3} & 0 & 1 & 0 & 0 & 0 & e^{i2\pi/3} & 0 \\ 0 & e^{-i2\pi/3} & 0 & 0 & 0 & 1 & 0 & e^{i2\pi/3} \\ e^{i2\pi/3} & 0 & 0 & 0 & 1 & 0 & e^{-i2\pi/3} & 0 \\ 0 & 0 & 0 & e^{-i2\pi/3} & 0 & e^{i2\pi/3} & 0 & 1 \\ 0 & 0 & e^{i2\pi/3} & 0 & e^{-i2\pi/3} & 0 & 1 & 0 \end{pmatrix} \begin{pmatrix} A_{11} \\ B_{11} \\ A_{12} \\ B_{12} \\ A_{21} \\ B_{21} \\ A_{22} \\ B_{22} \end{pmatrix} = E \begin{pmatrix} A_{11} \\ B_{11} \\ A_{12} \\ B_{12} \\ A_{21} \\ B_{21} \\ A_{22} \\ B_{22} \end{pmatrix}$$

(2)

where $A_{ij}$ and $B_{ij}$ are respectively the coefficients of $p_z^A$ and $p_z^B$ orbitals at the position $i \cdot \vec{a}_1 \times j \cdot \vec{a}_2$ in the linear combinations of the eigenfunctions (see Fig. 1c for the labelling of the C atom sites).

Obviously, the band gap at the Dirac point has to be (and it is) independent of the choice of the $m \cdot \vec{a}_1 \times n \cdot \vec{a}_2$ graphene SC. For any size of the SC the highest occupied level (HOL) and the lowest unoccupied one (LUL) are degenerate, with zero energy. HOL corresponds to the linear combination with all $A_{ij} = 0$ and all $B_{ij} = 1$. LUL corresponds to the linear combination with all the coefficients $A_{ij} = 1$ and all the coefficients $B_{ij} = 0$ (since the levels are degenerate, this correspondence is arbitrary and could be the opposite). However, if the size of the SC is $m \cdot \vec{a}_1 \times n \cdot \vec{a}_2$, being *both* $m$ and $n$ multiple of 3 (in the following this sort of SCs will be referred as 3C-SCs) a new feature arises: the second highest occupied level (HOL-1) and the second lowest unoccupied one (LUL+1) are also degenerate with the HOL and the LUL. This is because both inequivalent Dirac points, K and K', of the graphene Brillouin zone are coincident with the Γ point of a 3C-SC reciprocal cell. Thus, the solutions of the TB Hamiltonian for a 3C-SC at the K point also include the solutions at the K' point. For this reason there are four degenerate states with energy 0 for a 3C-SC, instead of the two found for the non-3C-SCs. In the case of non-3C-SCs the positions of K and K' points remain inequivalent. This feature can be better understood by looking at Fig.2, where the reciprocal lattices of a $3 \cdot \vec{a}_1 \times 3 \cdot \vec{a}_2$ SC (a 3C-SC) and a $4 \cdot \vec{a}_1 \times 4 \cdot \vec{a}_2$ SC (a non-3C-SC) are shown.

In the case of a 3C-SC the HOL and the LUL are the same as in the case of a non 3C-SC. However, since HOL and LUL are degenerate with the HOL-1 and the LUL+1 in a 3C-SC, it is possible to form new linear combinations of them that are also eigenfunctions of the Hamiltonian. For convenience, the following orthogonal linear combinations will be taken:

$$\begin{aligned} \text{HOL-1} &\rightarrow \text{All } B_{ij} = 0; A_{ij}^3 = 1; A_{ij}^2 = 0; A_{ij}^1 = e^{i\pi/3} \\ \text{HOL} &\rightarrow \text{All } A_{ij} = 0; B_{ij}^3 = 1; B_{ij}^2 = 0; B_{ij}^1 = e^{i\pi/3} \\ \text{LUL} &\rightarrow \text{All } A_{ij} = 0; B_{ij}^3 = 0; B_{ij}^2 = 1; B_{ij}^1 = e^{-i\pi/3} \\ \text{LUL+1} &\rightarrow \text{All } B_{ij} = 0; A_{ij}^3 = 0; A_{ij}^2 = 1; A_{ij}^1 = e^{-i\pi/3} \end{aligned} \quad (3)$$

where the superscripts 3, 2 and 1 in the $A_{ij}$ and $B_{ij}$ coefficients mean that $(i-j)$, $(i-j-2)$ and $(i-j-1)$ are multiple of 3, respectively.

It was already proved by Duplock et al. that the opening in the band gap for a $4\vec{a}_1 \times 4\vec{a}_2$ SC doped with a hydrogen atom bonded to one of the carbon atoms of the SC is caused by the potential of the H$^+$ ionic core[14]. Following the same idea it is straightforward to include the adsorption of a H atom in a $m \cdot \vec{a}_1 \times n \cdot \vec{a}_2$ SC in the TB model, just by adding a potential energy, $V$, to one of the diagonal elements of the TB matrix. Here, $V$ will be added to the first diagonal element of the matrix (corresponding to the $A_{11}$ site). In a non-3C-SC the HOL of the pristine graphene remains being an eigenfunction of the doped system, since the coefficient $A_{11} = 0$. By contrast, the LUL of the pristine graphene is not anymore an eigenfunction of the doped system, since the coefficient $A_{11} = 1$. Consequently the HOL and the LUL are not anymore degenerate in a non-3C-SC and a gap is opened at the Dirac point. This result of the TB model is confirmed by the DFT calculations of Duplock et al. for a $4\vec{a}_1 \times 4\vec{a}_2$ SC, where they found an opening in the gap of 1.25 eV[14]. The TB model also predicts the size of the gap between the HOL and the LUL in a non-3C-SC to be proportional to $\frac{1}{m \cdot n}$, i.e. the gap is proportional to the hydrogen concentration. This result can be checked using the MATLAB code given in the Supplementary Material[20].

In a 3C-SC the HOL, LUL and LUL+1 shown in Eq. (3) have a coefficient $A_{11} = 0$ and hence they remain being eigenfunctions of the TB Hamiltonian after the H atom adsorption. The HOL-1 has a coefficient $A_{11} = 1$ and thus is not anymore an eigenfunction after the adsorption. However, since three of the four frontier levels in a 3C-SC remain unaffected after the H adsorption it can be concluded that no gap is opened at the Dirac point after the adsorption of one H atom in a 3C-SC. Moreover, since all the coefficients $A_{ij}^3$ are equal to zero for the HOL-1, HOL and LUL, the following more general conclusion can be reached: *it is possible to adsorb several H atoms in a 3C-SC without a band gap opening at the Dirac point if all the H atoms are adsorbed at* $A_{ij}$ *sites, being* (i-j) *multiple of 3*. Henceforth this sort of arrangement will be called NGPs (non-gap periodicities). Given that $V$ is an adjustable parameter dependent on the kind of adsorbate, this result is also valid for other adsorbates different from atomic hydrogen.

It should be mentioned that Zhang et al.[21], through the Born approximation for pointlike scattering potentials, and Garcia-Lastra et al.[22], by means of the Non equilibrium Green's function formalism (NEGF), studied the ballistic transport of electrons in carbon nanotubes (CNTs). They found that for certain arrangements of two or more adsorbates on CNTs, which are particular cases of the NGPs presented here, there is an open channel close to the Fermi level for the electronic transport. That result is physically equivalent to the non band gap opening result shown in this work.

In order to validate the results of the TB model for the 3C-SCs, spin-polarized DFT calculations on $3\vec{a}_1 \times 3\vec{a}_2$, $6\vec{a}_1 \times 3\vec{a}_2$, $9\vec{a}_1 \times 3\vec{a}_2$, $12\vec{a}_1 \times 3\vec{a}_2$ and $6\vec{a}_1 \times 6\vec{a}_2$ SCs (all of them 3C-SCs) with a single H atom adsorbed were carried out. The calculations were performed by means of the PWSCF code[23]. The exchange-correlation energy was computed with the Perdew–Burke–Ernzehoff (PBE) functional[24]. Vanderbilt ultrasoft pseudopotentials replaced the ion cores. The kinetic energy cut-off was taken at 540 eV. The occupation of the one electron states was calculated at an electronic temperature of $k_BT = 0.01$ eV. The k-grid was built using the Monkhorst–Pack scheme[25], with maximum spacing between k points of 0.03 Å$^{-1}$.

The results of the DFT calculations are shown in table 1. The main difference between the DFT and the TB model results is that DFT includes the effect of the spin (i.e. the exchange energy) in the band gap. The HOL-1 of DFT calculations corresponds to a $p_z^A$ linear combination with a strong hybridization with the s orbital of hydrogen. This produces a significant decreasing in its energy (between 0.4 and 1.2 eV below the other three frontier levels in both spin channels, depending on the considered 3C-SC). The HOL, LUL and LUL+1 of DFT calculations can be identified with the ones obtained through the TB model (the ordering of the levels can vary due to exchange effects). In the spin up channel a tiny gap, varying from 0.19 eV in the $3\vec{a}_1 \times 3\vec{a}_2$ SC to 0.03 eV in the $12\vec{a}_1 \times 3\vec{a}_2$ SC, is opened between the other valence band and the two conduction bands. This is due to exchange effects not taken into account in the TB model. The effect of exchange is better observed in the density of states (DOS) of the doped graphene (Fig. 3). It can be noticed that close to the Fermi level the spin up channel electrons are stabilized with respect to the spin down ones because of the exchange effect. Obviously, the effect of exchange in the gap diminishes when the size of the unit cell increases. In the spin down channel the valence band and the two conduction bands are almost degenerate at the Dirac point, with a gap of 0.02 eV (for all of the studied 3C-SCs) between them. It is worth noticing that the gaps found in the studied 3C-SCs are one or two orders of magnitude smaller than the one found by Duplock et al. for the 4x4 SC (1.25 eV), showing that band gap opening in 3C-SCs are negligible in comparison to the non-3C-SCs ones. Furthermore this evidences the strong dependence of the band gap opening of doped graphene with the doping periodicity.

In conclusion, it has been shown that it is possible to adsorb atomic hydrogen in graphene without a band gap opening. No gap is opened in 3C-SCs, except for the small effect of exchange, when the H atoms are adsorbed at $A_{ij}$ sites, being (*i-j*) multiple of 3. This finding demonstrates that the gap in graphene can be tuned by controlling the positions of the adsorbed H atoms, opening up a new path for band gap engineering experiments. Moreover, these results are also valid for other adsorbates different from atomic hydrogen.

The author thanks Kristian S. Thygesen for useful discussions and Federico Calle-Vallejo for critical reading of the manuscript. This work was supported by the European Union through the FP7 e-I3 ETSF (Contract Number 211956), and THEMA (Contract Number 228539) projects. DFT calculations were carried out in the ARINA cluster.

**Tables**

**Table 1.** Energies (in eV) with respect to the Fermi level of the four frontier bands at the Dirac point for a graphene sheet with one hydrogen atom adsorbed with 3x3, 6x3, 9x3, 12x3 and 6x6 periodicities. S↑ and S↓ refer to spin up and down channels, respectively.

|  | 3x3 | | 6x3 | | 9x3 | | 12x3 | | 6x6 | |
|---|---|---|---|---|---|---|---|---|---|---|
|  | S↑ | S↓ | S↑ | S↓ | S↑ | S↓ | S↑ | S↓ | S↑ | S↓ |
| HOL-1 ($p_z^A$) | -1.77 | -1.72 | -1.11 | -1.08 | -1.14 | -1.15 | -0.71 | -0.70 | -0.78 | -0.77 |
| HOL ($p_z^B$) | -0.79 | -0.60 | -0.56 | -0.50 | -0.45 | -0.42 | -0.40 | -0.39 | -0.45 | -0.43 |
| LUL ($p_z^A$) | -0.60 | -0.62 | -0.47 | -0.48 | -0.40 | -0.40 | -0.37 | -0.37 | -0.41 | -0.41 |
| LUL+1 ($p_z^B$) | -0.79 | -0.60 | -0.56 | -0.50 | -0.45 | -0.42 | -0.40 | -0.39 | -0.45 | -0.43 |
| HOL-LUL gap | 0.19 | -0.02 | 0.09 | 0.02 | 0.05 | 0.02 | 0.03 | 0.02 | 0.04 | 0.02 |

**Figure captions**

**Figure 1.** (Color online) a) Basis vectors of the hexagonal lattice of graphene, $\vec{a}_{1,2} = \frac{a}{2}(\sqrt{3}, \pm 1)$, with $a = \sqrt{3} \cdot d_{C-C}$, being $d_{C-C}$ the carbon-carbon distance. Atoms in A (B) sites are in red (black). The coordinates of the A and B sites in the unit cell at the origin are (0,0) and $\frac{1}{3}(\vec{a}_1 + \vec{a}_2)$, respectively. $\vec{\delta}_1 = \left(\frac{a}{\sqrt{3}}, 0\right)$, $\vec{\delta}_2 = \left(-\frac{a}{2\sqrt{3}}, \frac{a}{2}\right)$ and $\vec{\delta}_3 = \left(-\frac{a}{2\sqrt{3}}, -\frac{a}{2}\right)$ are the nearest-neighbor vectors. (b) The first Brillouin zone of graphene with the K and K' points at $K',K = \left(0, \pm \frac{4\pi}{3a}\right)$. (c) A $3\cdot\vec{a}_1 \times 3\cdot\vec{a}_2$ supercell showing the criterion for the labeling of the C atoms followed in the present work.

**Figure 2.** (Color online) a) Brillouin zones of a 1x1 (black) and 3x3 (red) supercells of graphene. $\vec{b}_{1,2}$ and $\vec{b}^3_{1,2}$ are the reciprocal basis vectors of the 1x1 and 3x3 (red) supercells, respectively. High symmetry points of 1x1 Brillouin zone are in black without any superscript. The high symmetry points of 3x3 Brillouin zone are in red with a 3 superscript. The 3x3 supercell is also shown below. b) Brillouin zones of a 1x1 (black) and 4x4 (blue) supercells of graphene. $\vec{b}_{1,2}$ and $\vec{b}^4_{1,2}$ are the reciprocal basis vectors of the 1x1 and 4x4 (blue) supercells, respectively. High symmetry points of 1x1 Brillouin zone are in black without any superscript. The high symmetry points of 4x4 Brillouin zone are in blue with a 4 superscript. The 4x4 supercell is also shown below.

**Figure 3.** (Color online) Electronic density of states of a graphene sheet with a hydrogen atom adsorbed with a 3x3 periodicity. The blue curve corresponds to the spin up channel and the red one to the spin down channel. The energies are given with respect to the Fermi level.

**Figure 1**

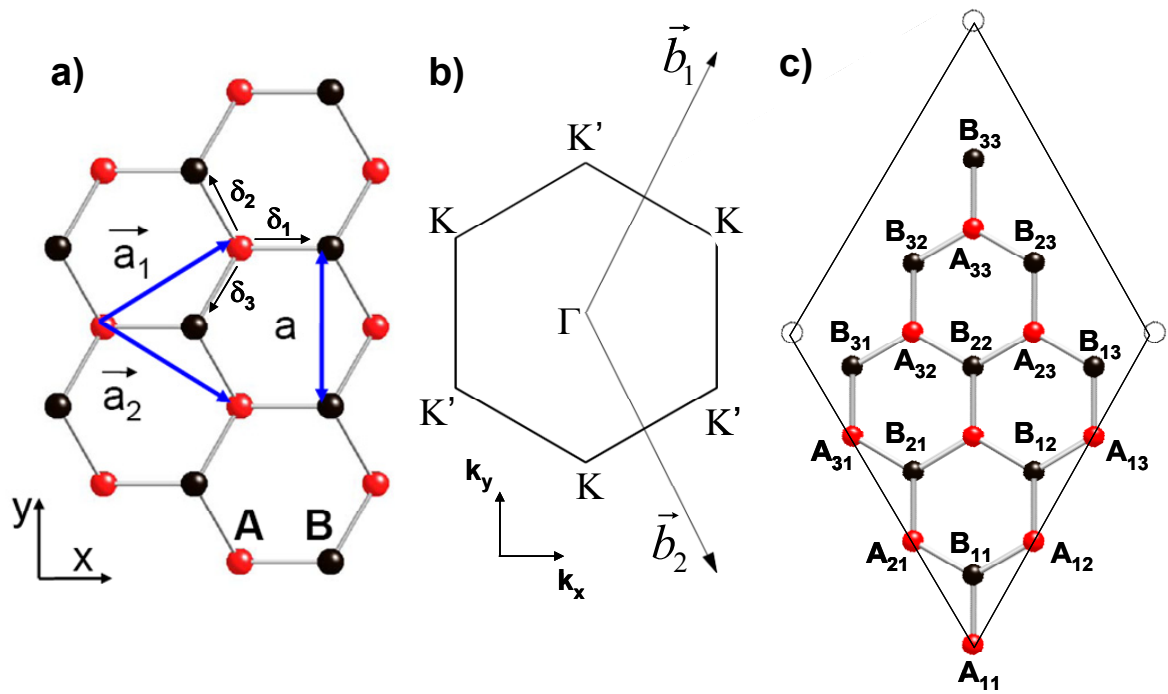

**Figure 2**

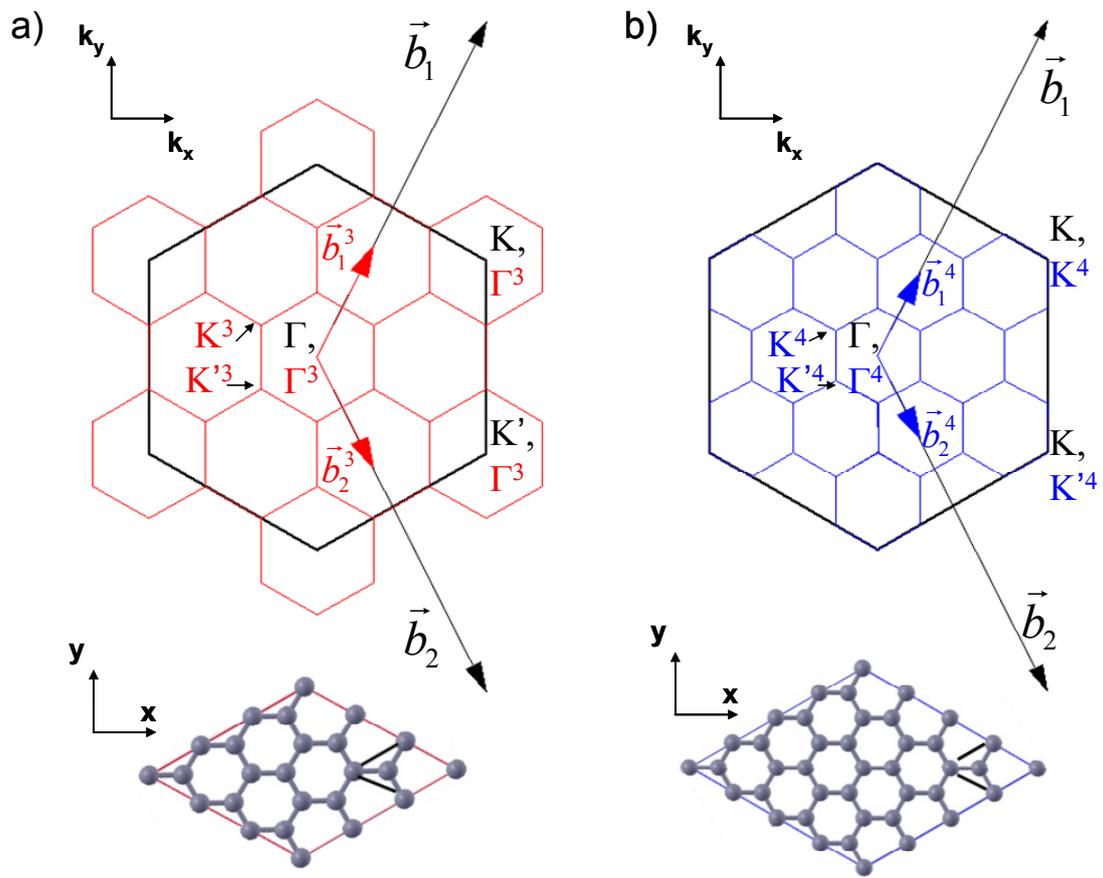

**Figure 3**

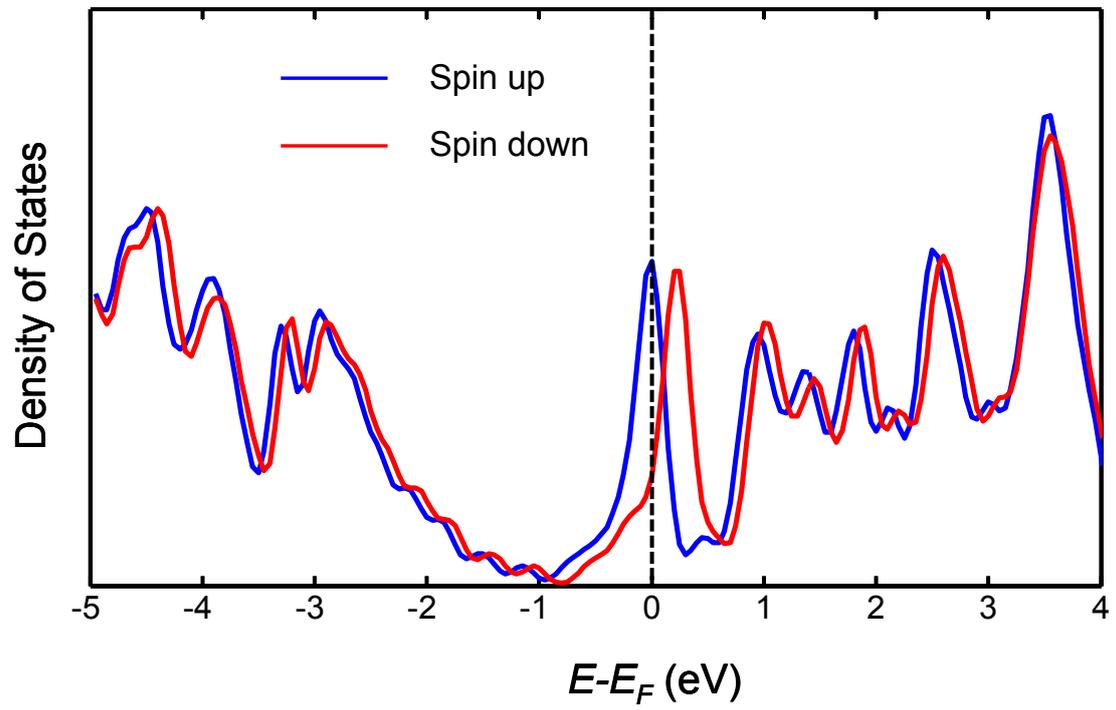

# Supplementary Material for

# "Adsorption of Hydrogen in Graphene without Band Gap Opening at the Dirac Point"

by

Juan María García-Lastra

**Tight-binding model results**

Figure 1 shows the dependence of the band gap opening of doped graphene in the tight-binding model as a function of the size of the unit cell. It can be seen that the band gap opening is zero when both *m* and *n* are multiple of 3. When this is not the case it can be noticed that the gap decays as $1/(m \cdot n)$. The MATLAB script used along this work is shown below.

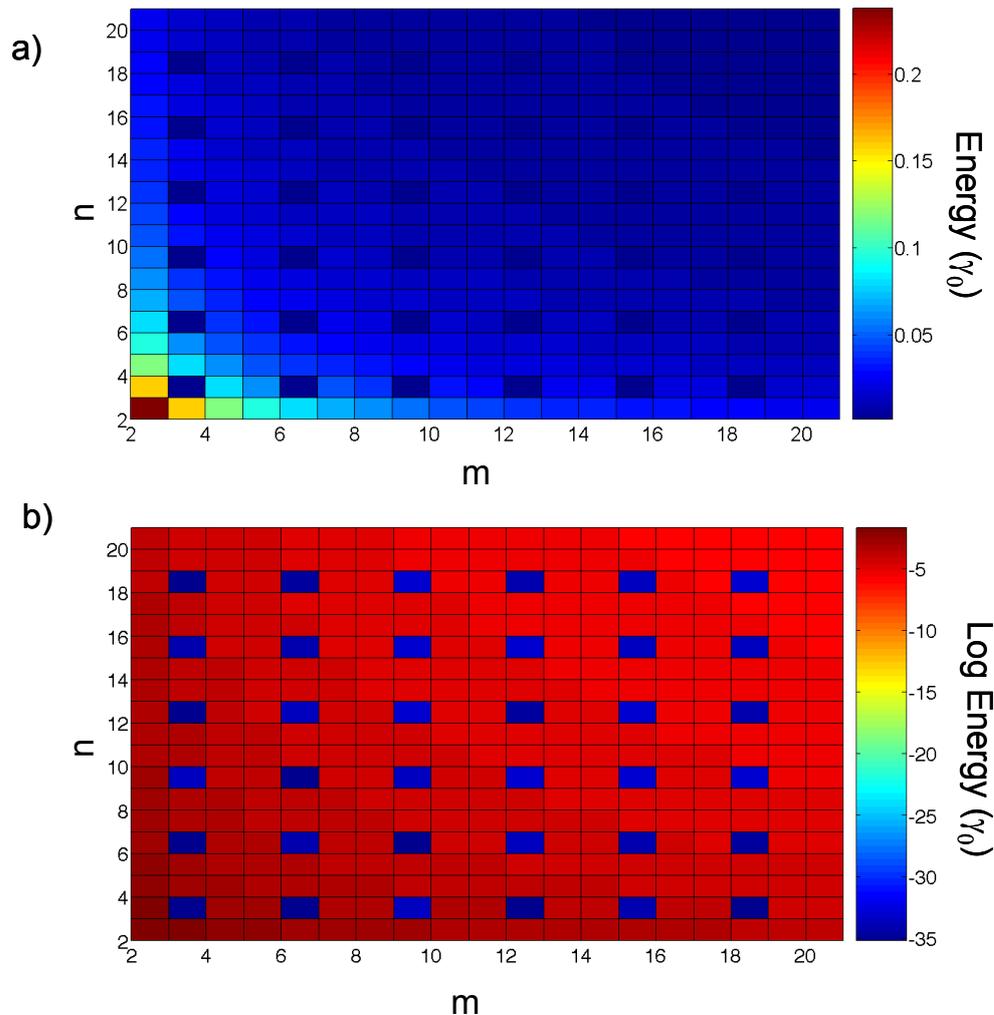

**Figure 1.** Band gap opening as a function of the size of the unit cell (nxm) for a value of the potential at $A_{11}$ site of V= $\gamma_0$ in a linear scale (a) and in a logarithmic scale (b) for the energies. Energy in units of $\gamma_0$ (the energy of the hopping integral between the $p_z$ orbitals of two adjacent carbon atoms in a CNT).

```matlab
clear
% m and n are the size of the supercell (ma1 x na2). The script is built in
% such a way that m =< n
m=3;
n=6;
if m>n
    p=m;
    m=n;
    n=p;
end
% val is the value of the potential at A11 site
val=1.0;
% fas is a particular complex number
fas=exp(i*pi/3);
% mat is the matrix of the TB model
mat=zeros(2*m*n,2*m*n);
% comp1, comp2, comp3 and comp4 are the HOL-1, HOL, LUL and LUL+1 of Eq.
% 1.3. They will be used when m and n are multiple of 3. We do this because
% Matlab shows random linear combinations of them as eigenvectors.
% To check thatthey are solutions do mat*comp.
if mod(m,3)==0 & mod(n,3)==0
    comp1=zeros(2*m*n,1);comp2=comp1;comp3=comp1;comp4=comp1;
end
for j=1:m*n
    % A sites have odd indexes in the matrix. B sites have even
    % indexes. x_j and y_j are the i and j indexes of the Carbon sites
    % following the figure 1c of the manuscript. x_1, y_1, x_2, y_2, x_3
    % and y_3 and the i and j indexes of the first neighbours carbon sites.
    ind_j=2*j-1;
    x_j=fix((j-1)/n)+1;
    y_j=j-(x_j-1)*n;
    x_1=x_j;
    y_1=y_j;
    x_2=x_j;
    y_2=y_j-1;
    if y_2==0
        y_2=n;
    end
    x_3=x_j-1;
    if x_3==0
        x_3=m;
    end
    y_3=y_j;
    pos_1=n*(x_1-1)+y_1;
    pos_2=n*(x_2-1)+y_2;
    pos_3=n*(x_3-1)+y_3;
    ind_1=pos_1*2;
    ind_2=pos_2*2;
    ind_3=pos_3*2;
    mat(ind_j,ind_1)=mat(ind_j,ind_1)+1;
    mat(ind_1,ind_j)=mat(ind_1,ind_j)+1;
    mat(ind_j,ind_2)=mat(ind_j,ind_2)+exp(i*2*pi/3);
    mat(ind_2,ind_j)=mat(ind_2,ind_j)+exp(-i*2*pi/3);
    mat(ind_j,ind_3)=mat(ind_j,ind_3)+exp(-i*2*pi/3);
    mat(ind_3,ind_j)=mat(ind_3,ind_j)+exp(i*2*pi/3);
    % This is the part used to build HOL-1, HOL, LUL and LUL+1 when m and n
```

```matlab
        % are multiple of 3
        if mod(m,3)==0 & mod(n,3)==0
            if mod(x_j-y_j,3)==0
                comp1(ind_j)=1;
                comp1(ind_j+1)=0;
                comp2(ind_j)=0;
                comp2(ind_j+1)=0;
                comp3(ind_j)=0;
                comp3(ind_j+1)=1;
                comp4(ind_j)=0;
                comp4(ind_j+1)=0;
            end
            if mod(x_j-y_j,3)==1
                comp1(ind_j)=fas;
                comp1(ind_j+1)=0;
                comp2(ind_j)=fas';
                comp2(ind_j+1)=0;
                comp3(ind_j)=0;
                comp3(ind_j+1)=fas;
                comp4(ind_j)=0;
                comp4(ind_j+1)=fas';
            end
            if mod(x_j-y_j,3)==2
                comp1(ind_j)=0;
                comp1(ind_j+1)=0;
                comp2(ind_j)=1;
                comp2(ind_j+1)=0;
                comp3(ind_j)=0;
                comp3(ind_j+1)=0;
                comp4(ind_j)=0;
                comp4(ind_j+1)=1;
            end
        end
end
% We put the potential V at the A11 site. We can also put a second
% potential V at other site.
mat(1,1)=val;
% The matrix is diagonalized.
[b,c]=eig(mat);
[a,d]=sort(diag(c));
% The gap between the HOL and the LUL is obtained
gap=a(n*m+1)-a(n*m);
```